\documentclass[aps,11pt,axodraw,nofootinbib,superscriptaddress,aps]{revtex4}
\pdfoutput=1
\usepackage{epsfig}
\usepackage{amsmath}
\usepackage{bm}
\usepackage{times}
\usepackage{graphicx}
\usepackage{epstopdf}
\usepackage{amsfonts}
\usepackage{bm}
\usepackage{epsfig}
\usepackage{graphics}
\usepackage{xspace}
\usepackage[usenames]{color}
\usepackage{chngpage}

\def\nn{\nonumber}
\def\bea{\begin{eqnarray}}
\def\eea{\end{eqnarray}}
\def\ba{\begin{eqnarray}}
\def\ea{\end{eqnarray}}
\def\be{\begin{equation}}
\def\ee{\end{equation}}

\def\beq{\begin{equation}}
\def\eeq{\end{equation}}

\def\nn{\nonumber}

\begin{document}

\title{\Large Hitting sbottom in natural SUSY}

\author{Hyun Min Lee}
\affiliation{Theory Division, Physics Department, CERN, CH-1211 Geneva 23, Switzerland}

\author{Veronica Sanz}
\affiliation{Theory Division, Physics Department, CERN, CH-1211 Geneva 23, Switzerland}
\affiliation{Department of Physics and Astronomy, York University, Toronto, ON, Canada, M3J 1P3}

\author{Michael Trott}
\affiliation{Theory Division, Physics Department, CERN, CH-1211 Geneva 23, Switzerland}

\date{\today}
\begin{abstract}
We compare the experimental prospects of direct stop and sbottom pair production searches at the LHC.
Such searches for stops are of great interest as they directly probe for states that are motivated by the SUSY solution to the hierarchy
problem of the Higgs mass parameter - leading to a ``Natural" SUSY spectrum. Noting that sbottom searches are less experimentally challenging and scale up in reach directly with the improvement on b-tagging algorithms, 
we discuss the interplay of small TeV scale custodial symmetry violation with sbottom direct pair production searches as a
path to obtaining strong sub-TeV constraints on stops in a natural SUSY scenario. We argue that if a weak scale natural SUSY spectrum does not exist within the reach of LHC,  then hopes for such a spectrum for large regions of parameter space should ``sbottom out". Conversely, the same arguments make clear that a discovery of such a spectrum is likely to proceed in a ``sbottom up" manner.
\end{abstract}
\maketitle
\section{Introduction.}

Supersymmetry (SUSY)\footnote{See Refs. \cite{review,Martin:1997ns} for reviews.} has been the leading paradigm of physics Beyond the Standard Model (BSM) for the last three decades, and searching for SUSY is the leading BSM priority of the experimental particle physics community. Unfortunately, despite extensive and continued searches at generations of colliders and underground detectors, no direct evidence of weak scale SUSY particles have been found to date. LEP, and now the Large Hadron Collider (LHC) has pushed the direct production bounds on SUSY particles  away from the natural expectation for their mass scale, proximate to the weak scale, with the bounds on SUSY colored particles now in the TeV range in large regions of parameter space.
The interpretation of experimental results in missing transverse energy ($E_T \! \! \! \! \! \! \! /$ \,\,) searches in terms of the scale of SUSY particles is model dependent, but it is difficult (although clearly not impossible\footnote{See Refs.~\cite{Essig:2011qg,Kats:2011qh,Brust:2011tb,Papucci:2011wy,naturalsusy,Csaki:2012fh} for some discussion of SUSY scenarios consistent with the most recent experimental bounds. In order to reduce the experimental profile of a SUSY spectrum, to be compatible with the non observation of superparticles to date, a minimal effective SUSY spectrum which is composed only of the third generation scalar superpartners, gauginos and Higgsinos is generally entertained. See also Refs. \cite{Dimopoulos:1995mi,Cohen:1996vb,Csaki:2011ge}.})
to design models which compellingly explain the basic experimental result; namely, that SUSY partners have not been found, to date, residing at mass scales where they naturally should be.

The question of whether a SUSY theory is natural, in that it avoids excessive fine tuning, is largely tied to the mass scale of third generation sfermion partners to the SM fermions with the largest couplings to the electroweak symmetry breaking (EWSB) sector.
The couplings of the third generation sfermions transmit the soft SUSY breaking mass scale $\rm M_{SUSY}$ to this sector, which leads to large perturbative corrections to Lagrangian parameters linked to the EWSB scale $v \sim 246 \, {\rm GeV}$ without fine tuning.\footnote{For a recent discussion on fine-tuning in traditional GUT-based SUSY models in light of the recent LHC results, see Ref.~\cite{finetune}.} 
 

The question of naturalness when considering natural SUSY sfermion spectra has largely focused on the bounds on the mass scale of stops to date, and in particular direct production bounds on stops.\footnote{See however Ref. \cite{Papucci:2011wy} for a rather comprehensive summary of the current experimental limits on stop and sbottoms.}
However, missing energy based signal isolation is not as powerful for such searches, as the stop typically decays into final states too similar to SM top production backgrounds which also has a missing energy component due to the decay $t \rightarrow b \, \bar{\nu} \, \ell$. Overcoming SM backgrounds is a major challenge in stop searches, which also do not scale well with the invariant mass of the stop due to the complex decay topology, as we will discuss.\footnote{See Ref.~\cite{Bai:2012gs} for an attempt to tame the large SM background associated with the generically difficult stop signal.}
Indeed, whereas broad bounds on first and second generation squarks exist  \cite{Chatrchyan:2011zy,Aad:2011ib}, and on sbottoms up to $390 \, {\rm GeV}$ at $95 \%$ CL for neutralino masses below $60 \, {\rm GeV}$ \cite{direct-sbottom}, generic searches of this form for directly produced stops decaying into $t \, \bar{t}$+$E_T \! \! \! \! \! \! \! /$ \,\, are not sensitive to the cross sections expected for stops to date.\footnote{See Refs.~{\cite{Abazov:2009ps,Aaltonen:2011rr,Aaltonen:2011na,Aad:2011wc}} for searches related to direct stop production.} This is due to a combination of signal efficiency and $t \, \bar{t}$ background contamination constraints. 
   
However, one can relate sbottom searches to the stop sector, as the splitting of the left-handed components of stops and sbottoms are bounded by precision measurements of the $W$ boson's properties at LEP. Hence, as the sbottom searches become more sensitive, the bound on the stop sector becomes stronger through an interplay of direct sbottom limits and Electroweak Precision Data (EWPD). A relationship of this form between sbottom and stop exclusion regions also follows from the fact that soft SUSY masses are invariant under $\rm SU_L(2)$. In this way, sbottom direct search limits can strongly drive stop limits in a manner that can provide more experimental reach than direct stop searches alone, particularly in the regime of stop masses most of interest in natural SUSY. As a result, there is a tension between naturalness, which limits the stop scale from above, and sbottom searches, which restrict the stop sector from below, which might become acute in the 2012 data set. 

The paper is outlined as follows.
In Section \ref{theory} we discuss the strong theoretical link between the sbottom and stop masses due to theoretical consistency and EWPD, and also discuss how naturalness concerns
influence the expected stop masses in combination with Higgs mass constraint. In Section \ref{direct} we will discuss prospects for exclusion limits on direct pair production
of sbottoms and stops. In Section \ref{interplay} we demonstrate the interplay of indirect constraints on stop masses and show how this leads to our conclusions.

\section{Minimal Consistency Constraints.}\label{theory}

In this section we consider three sources of minimal consistency constraints on the stop sector before discussing the experimental prospects of stop and sbottom searches in the following section. The first constraint is associated with the splitting between stops and sbottoms and the contribution to the $\Delta \rho$ parameter, which restricts the amount of custodial symmetry -- $\rm SU_C(2)$-- violation. The second restriction comes from the paradigm of natural SUSY with minimal fine tuning, namely that the stop sector should be close to the electroweak symmetry breaking scale. Finally, one can draw further conclusions on the stop sector by looking at the Higgs mass in the minimal SUSY model (MSSM). This last requirement is more model dependent, as the Higgs may receive contributions to lift its mass in non minimal SUSY models - such as from an SM singlet as in the next to minimal SUSY model (NMSSM) \cite{Fayet:1974pd,Dine:1981rt}.

\subsection{$\rm SU_C(2)$ Violation}

In a natural SUSY model, the squarks are characterized by the following mass matrix
\bea
\mathcal{L}_{m_{\tilde{f}}} = - \frac{1}{2} \left(\tilde{f}^\dagger_L , \tilde{f}^\dagger_R \right) \, {\bf \mathcal{Z}}  \left(
\begin{array}{c} 
\tilde{f}_L \\ 
\tilde{f}_R \\
\end{array} \right),
\eea
where
\bea
{\bf \mathcal{Z}}  = \left(\begin{array}{cc} 
\cos^2 \theta_{\tilde{f}} \, m_{\tilde{f}_1}^2 + \sin^2 \theta_{\tilde{f}} \, m_{\tilde{f}_2}^2 & \quad  \, \sin \theta_{\tilde{f}} \cos \theta_{\tilde{f}} \left(m_{\tilde{f}_1}^2 - m_{\tilde{f}_2}^2 \right)\\ 
\sin \theta_{\tilde{f}} \cos \theta_{\tilde{f}} \left(m_{\tilde{f}_1}^2 - m_{\tilde{f}_2}^2 \right) & \quad \sin^2 \theta_{\tilde{f}} \, m_{\tilde{f}_1}^2 + \cos^2 \theta_{\tilde{f}} \, m_{\tilde{f}_2}^2 \\
\end{array} \right).
\eea
The cosine ($c_{\tilde{f}}$) and sine ($s_{\tilde{f}}$) of the squark mixing angles $\theta_{\tilde{f}}$ and the physical masses $m_{\tilde{f}_1},m_{\tilde{f}_2}$ are derivable from the initial soft SUSY breaking chiral squark masses 
in the Lagrangian corresponding to the fields $\tilde{f}_{L/R}$. We also define a mass difference for later convenience $\delta m^2 = m_{\tilde{t}_2}^2 - m_{\tilde{t}_1}^2>0$. The  off-diagonal entry in the squark mass matrix is proportional to the corresponding fermion mass, as such we will neglect sbottom mixing,
implicitly restricting ourselves to a moderate value of $\tan \beta$ regime in this paper. 
Explicitly, our convention is that $m_{\tilde{f}_1}$ corresponds to the left handed sfermion in the limit of
no mixing, and this is the limit we adopt for the sbottom states. This choice is conservative, as we will relate the experimental bound on the lightest sbottom to the stop sector. If the lightest sbottom experimentally bounded was purely right-handed, the bounds that we will discuss would actually be stronger, as the left handed sbottom would then be heavier, enforcing stronger (although unquantified) bounds on the stop sector. This is true so long as the spectrum is such that left handed
sbottoms are not experimentally inaccessible compared to right handed sbottoms. We will quantify this condition on the sbottom branching ratio in what follows.

In the MSSM, a relationship is enforced between sbottom and stop direct search bounds because soft SUSY masses are $\rm SU_L(2)$ invariant\footnote{The mass relation is still true of the non-minimal SUSY models if there is no additional EWSB other than the MSSM Higgs sector.}. At leading order, this results in the well known relation
\bea
m_{\tilde{b}_1}^2 \approx  \cos^2 \theta_{\tilde{t}} \, m_{\tilde{t}_1}^2 + \sin^2 \theta_{\tilde{t}} \, m_{\tilde{t}_2}^2 - m_t^2 - m_W^2 \, \cos (2 \, \beta).
\label{softSU2}
\eea
Here we have considered small sbottom mixing, $\cos^2 \theta_{\tilde{b}} \sim 1$. We will neglect perturbative corrections to this relationship. This relation does not tie the sbottom limit to the lightest stop ($\tilde{t}_1$) necessarily, but only to the left-handed composition -- $\tilde{t}_L$. In the worst case scenario, where the lightest stop would be purely right-handed $m_{\tilde{t}_2}^2 >m_{\tilde{b}, min}^2 +m_t^2$, and there is no direct prediction on the lightest stop from this relation. In the next section we will show how naturalness in minimal models-- a key motivation of natural SUSY -- changes this picture, and can lead to stronger conclusions as it selects for a nearly degenerate stop spectrum.

Besides this relation, there is a simple interplay in a
minimal sfermion spectrum between EWPD and direct sbottom and stop production searches. Limits from EWPD quantifies the bounds on non SM interactions that modify the vacuum polarizations of the $W^\pm, Z$ bosons,
characterized by the $\rm STU$ parameters \cite{Holdom:1990tc,Peskin:1991sw,Altarelli:1990zd}. Fits to these parameters can be re-interpreted if the Higgs hints at $\sim 125 \, {\rm GeV}$ are confirmed. With a Higgs mass fixed to this prior value, 
EWPD then gives a direct constraint (or direct measure) on $\rm SU_C(2)$ breaking physics in a natural SUSY sfermion spectrum.
For EWPD fits, we use the results of the ${\it Gfitter}$ \cite{Baak:2011ze}
\bea
S = 0.02 \pm 0.11,  \quad \quad  T = 0.05 \pm 0.12,  \quad \quad U =
0.07 \pm 0.12 \;.
\eea
We include a correction to $\rm STU$ of the form $(\Delta S, \Delta T, \Delta U)_{mh = 125} = (0.004,-0.003,-0.0001)$ due to shifting the best fit value of the Higgs mass in these
fit results from $120$~GeV to $125 \, {\rm GeV}$ using the one-loop Higgs boson contribution to $\rm STU$. The relevant quantity 
for constraining non SM $\rm SU_C(2)$ violation is
\bea
(\Delta \rho_0)^{\pm}_L &=& (\rho_0)_{mh = 125}  - 1, \nn \\ 
&=&  \hat{\alpha}(m_z) \, (T +  (\Delta T)_h), \\
&=&(3.67 \pm 8.82) \times 10^{-4}.
\eea
Here we have taken the PDG value $\hat{\alpha}(m_z) = 127.916 \pm 0.015$.
The dominant one loop contribution to this quantity in natural SUSY spectra arises from the one loop scalar top and bottom contribution.
Explicitly it is given by the following expression \cite{Barbieri:1983wy,Drees:1990dx,Chankowski:1993eu,Heinemeyer:2004gx} where we neglect terms proportional to small sbottom mixing angles
\begin{align}\label{deltarho}
\Delta \rho_0^{SUSY} &\approx  \frac{3\, G_F \,  \cos^2 \theta_{\tilde{t}}}{8 \, \sqrt{2} \, \pi^2} \, \left\{- \sin^2 \theta_{\tilde{t}} \, F_0[m^2_{\tilde{t}_1}, m^2_{\tilde{t}_2}]
+ F_0[m^2_{\tilde{t}_1}, m^2_{\tilde{b}_1}] +\tan^2 \theta_{\tilde{t}} \, F_0[m^2_{\tilde{t}_2}, m^2_{\tilde{b}_1}] \right\}.
\end{align}
The function $F_0$ is defined as
\bea
F_0[x, y] = x + y - \frac{2 \, x \, y}{x - y} \, \log \frac{x}{y}.
\eea
It is instructive to consider the constraint $\Delta \rho_0^{SUSY}  \lesssim (\Delta \rho_0)^+_L$. We show this constraint in  Fig. (1) when a lower bound on $m_{\tilde{b}_1}$ is fixed to various values.
\begin{figure}[h!]
\includegraphics[scale=0.49]{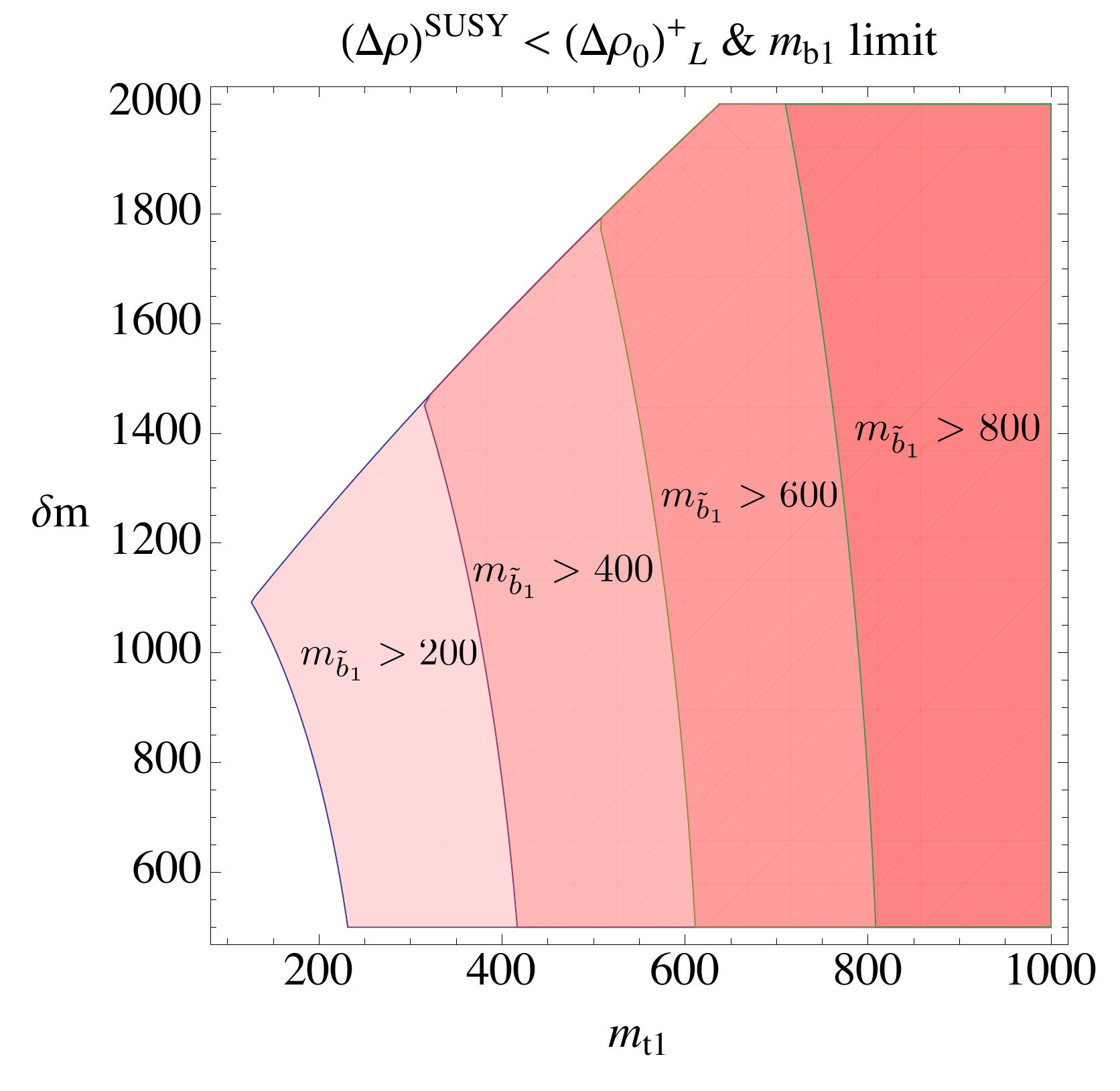}
\includegraphics[scale=0.49]{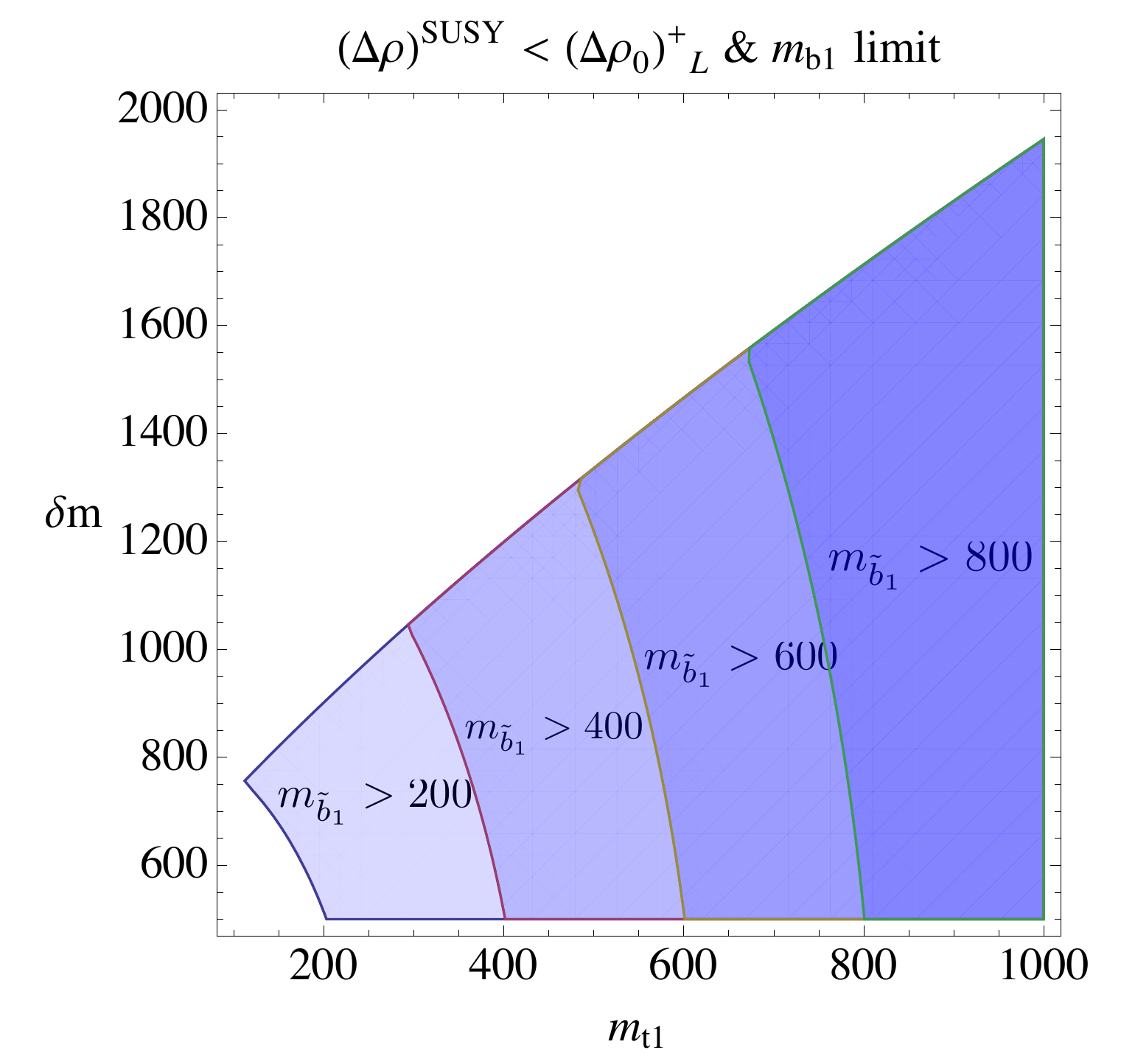}
\includegraphics[scale=0.49]{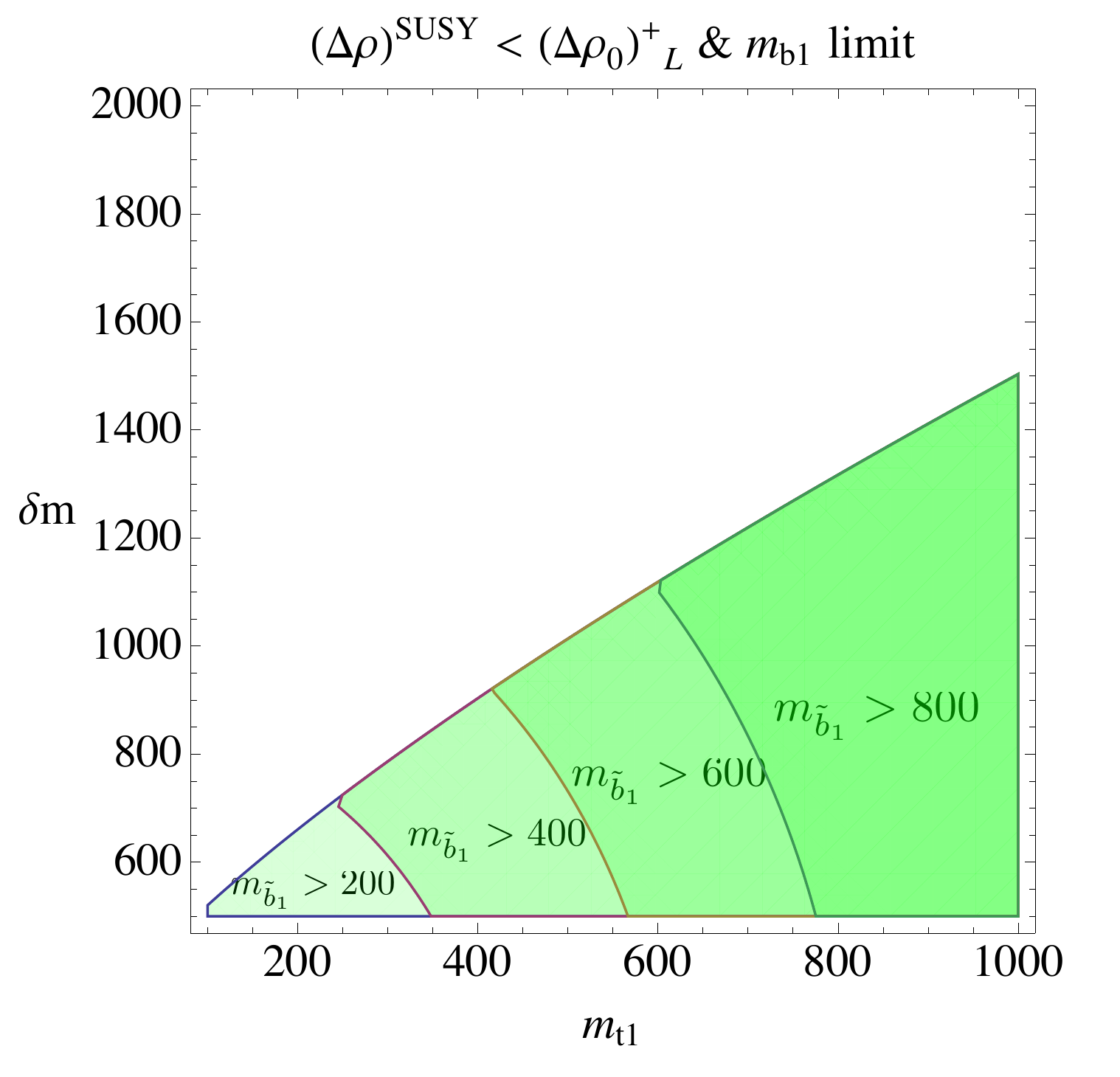}
\includegraphics[scale=0.51]{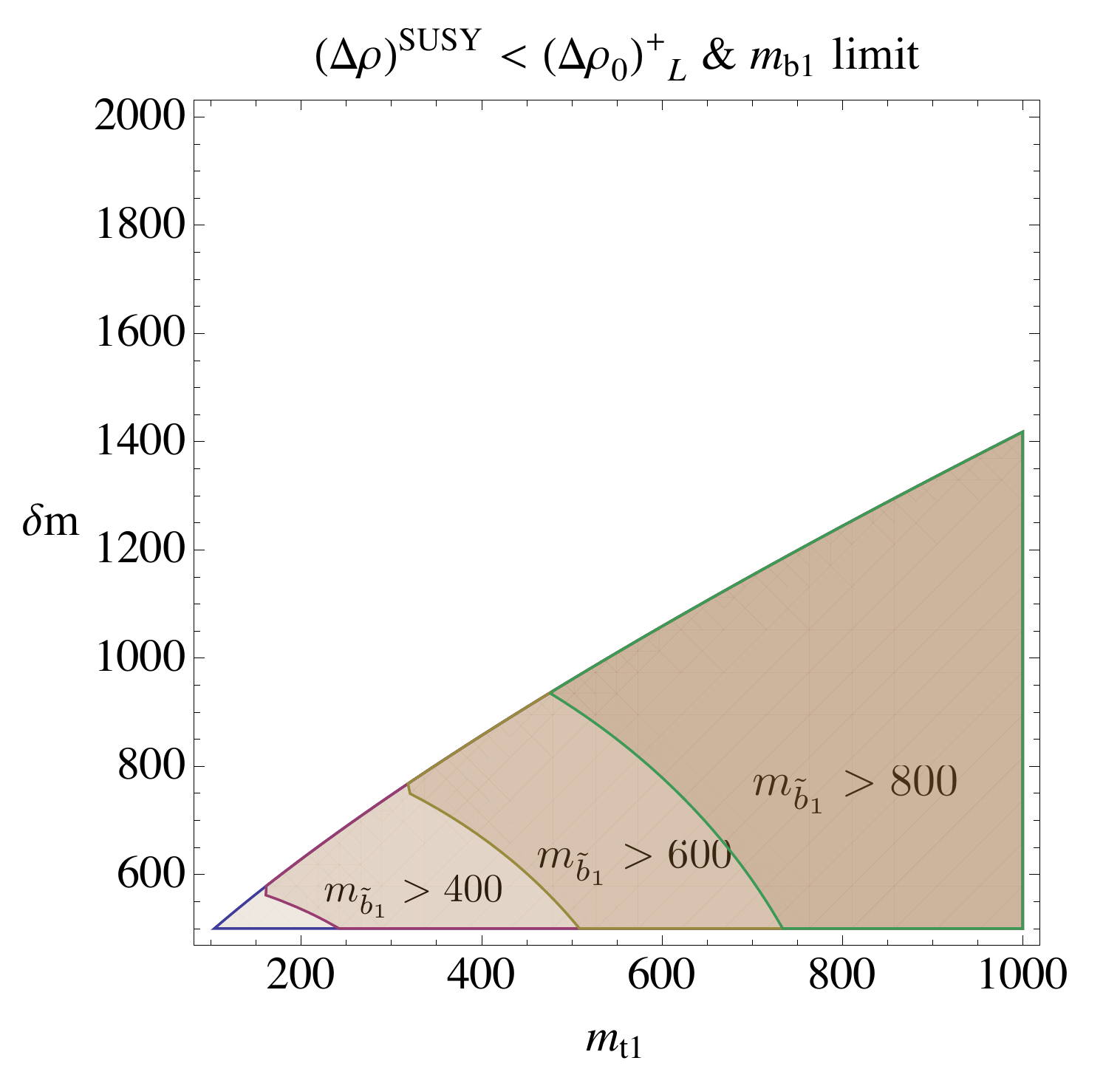}
\caption{The allowed $(m_{\tilde{t}_1}, \delta m)$ consistent with the $\Delta \rho$ constraint and an imposed lower bound on $m_{\tilde{b}_1}$, for various stop mixing angles. The colour coding of the plots
is systematic in this section.The top left figure, in red, corresponds to the allowed regions considering these constraints for 
$(\tan \beta, \sin \theta_t) = (10,0.2)$ and various $m_{\tilde{b}_1}$ lower bounds, the top right figure shows a set of blue regions with $(10,0.3)$, the bottom left figure shows a set of green regions where $(10,0.5)$ and the bottom right plot shows allowed regions in brown where $(10,1/\sqrt{2})$ - this last case corresponding to maximal mixing. These plots are not sensitive to $\tan \beta$.}
\end{figure}

Due to EWPD constraints and/or simply insisting on $\rm SU_L(2)$ preserving soft masses, raising the direct exclusion limit of $m_{\tilde{b}_1}$  
indirectly yields an exclusion constraint in the space of $(m_{\tilde{t}_1},m_{\tilde{t}_2}, \theta_{\tilde{t}})$. The constraints on this space are likely to be driven by sbottom
searches for large regions of parameter space, as we will discuss. This is fortunate for efficiently raising stop mass limits and addressing the question of when a natural SUSY paradigm for particular parameters in the
stop sector is experimentally ruled out. In this manner, the parameter space for such natural SUSY sfermion spectra can sbottom out experimentally. Conversely, these same arguments make clear that
a sbottom up discovery of the third generation sfermion spectra is a scenario experimentally favoured in large regions of parameter space.

\subsection{Natural SUSY and the stop/sbottom splitting}

A natural SUSY spectrum must also
confront theoretical consistency in the form of fine tuning considerations. 
Although it is difficult to define a uniquely compelling fine tuning measure, or argue what degree of fine tuning is clearly unacceptable, 
a popular fine tuning measure is based on the required cancelation of the tree level and loop contributions to the $Z$ boson mass.
An ominous level of fine tuning is widely considered to be $\sim 1\%$.
We use a fine tuning measure inspired by  \cite{Ellis:1986yg,Barbieri:1987fn}. The $Z$ mass can receive contributions from many SUSY breaking sources, but unavoidably, the stop sector contributes to the $Z$ via the stop contributions to the up-type Higgs, $\delta m_{H_u}$. Therefore, when using this measure, we will consider the theory to be at least fine-tuned by the splitting between stops and $m_Z$. This fine tuning measure is given by
\bea\label{deltaz}
\Delta_Z > \Delta_Z^t = \left| \frac{\delta_t \, m_Z^2}{m_Z^2} \right|.
\eea
Restricting ourselves to the moderate value of $\tan \beta$ and further assuming a degree of degeneracy in the mass spectrum of heavier squarks so that the impact of the stop loops remains largest, one finds \cite{Okada:1990vk,Ellis:1990nz,Haber:1990aw,Perelstein:2007nx} at one loop
\bea
\delta_t \, m_Z^2 = \frac{3}{16 \, \pi^2} \, \left(y_t^2 (m_{\tilde{t}_1}^2 + m_{\tilde{t}_2}^2 - 2 m_t^2) + \frac{(m_{\tilde{t}_1}^2 - m_{\tilde{t}_2}^2)^2}{4 \, v^2 \, \sin^2 \beta} 4 \, c_{\tilde{t}}^2 \, s_{\tilde{t}}^2 \right) \, \log \left(\frac{2 \, \Lambda^2}{m_{\tilde{t}_1}^2 + m_{\tilde{t}_2}^2} \right). \label{finetune}
\eea
Here $\Lambda$ is the scale associated with new states required to cut off the logarithmic divergence resulting form the splitting of the stop and top masses and is associated
with the messenger scale. For numerical purposes we conservatively consider the cut off scale to be taken to be a factor of 100 above the expected approximate geometric mean of the stop masses $ \sim  1 \, {\rm TeV}$. 
Note that the Eqn.~(\ref{finetune}) neglects $1/\tan^2 \beta$ corrections, so that our analysis is essentially restricted to a range, $2 \lesssim \tan \beta \lesssim 20$, when we impose this constraint. The upper limit follows from the assumed dominance of stop loops in Eqn.~(\ref{Higgsmass}) and can be relaxed. 
In a natural SUSY spectrum, one also expects light Higgsinos as their mass is driven by the $\mu$ parameter, and relatively light gluinos with a mass scale $\lesssim 1 \, {\rm TeV}$. This later expectation follows from the 
one loop correction that gluinos generate for stop masses, which contributes to Eqn.(\ref{deltaz}) at two loops. The contributions from these particles to this fine tuning measure are sub dominant and neglected. This is a conservative choice.

\begin{figure}[h!]
\centering
\includegraphics[scale=0.22]{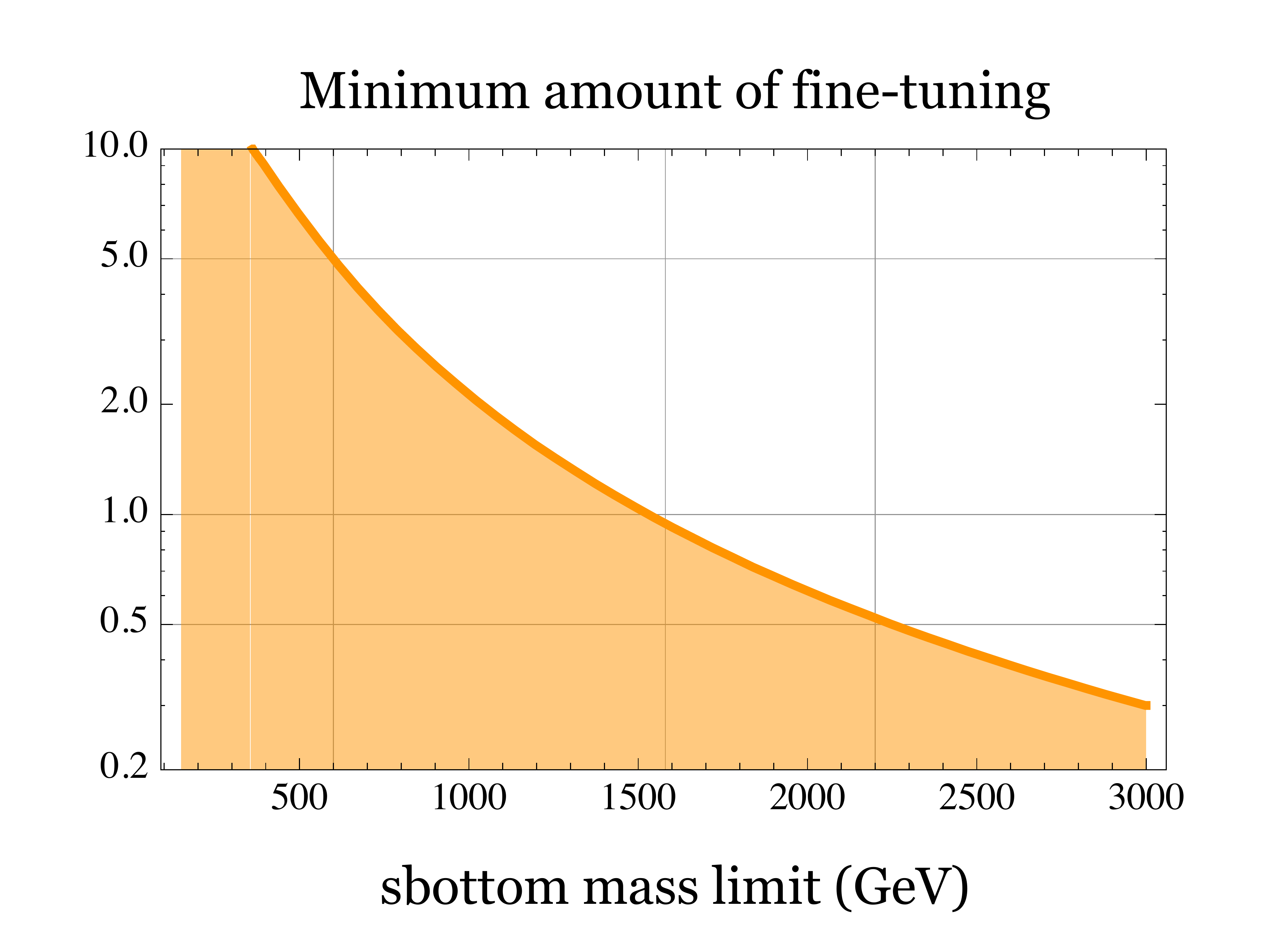}
\includegraphics[scale=0.22]{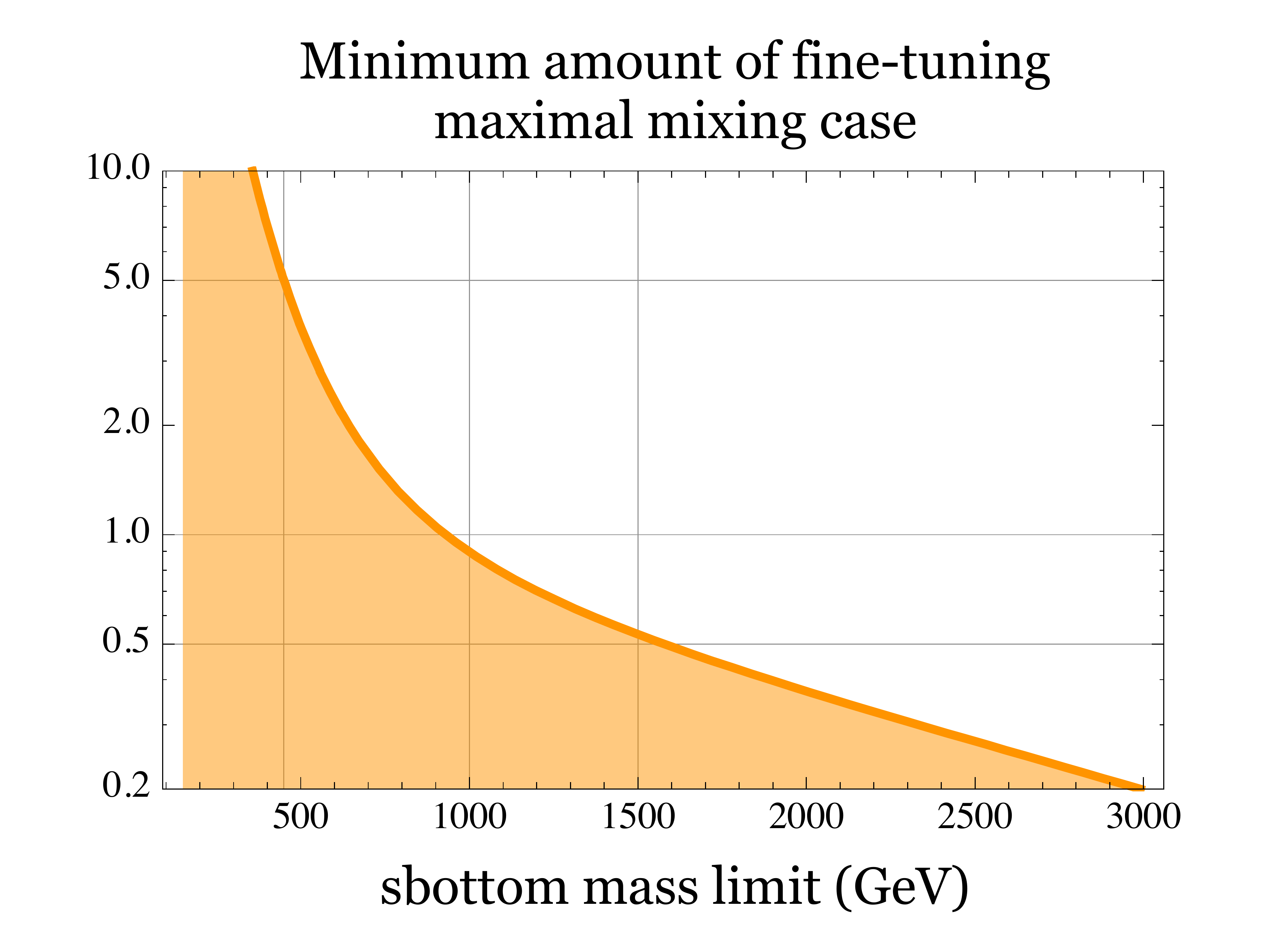}
\caption{Fine-tuning measure (in $\%$) for different bounds on sbottom particles. Left: general case. Right: maximal mixing case. Here $\tan\beta=10$ and $\Lambda=100$ TeV.}
\label{finetuning}
\end{figure}
In Fig.~(\ref{finetuning}) we show the maximum value of $\Delta_Z^t$ as a function of the sbottom limit, using the constraints from $\Delta \rho$ and relation Eq.~(\ref{softSU2}). We varied the stop parameters in a completely general way, (i.e. we do not impose the MSSM Higgs mass constraint here) and also show how the finetuning scales with the sbottom bounds faster as we approach maximal mixing. 
If the sbottom limit is increased to 500 GeV, one already knows that the tuning of the theory is {\it at least} at the level of 5\% according to this measure, whereas pushing the limits of sbottoms to $2 \, {\rm TeV}$ translates in an increase of the tuning to the 0.5\% region. 

Our approach is not to invoke any particular UV model dictating a full SUSY spectrum. And we note that our parameter choices, such as the numerical values of $\Lambda$, are conservative. 
However, the argument we advance can be made more precise at the cost of more assumptions in the UV structure of the theory. For example,  in minimal gauge mediation \cite{gaugemediation}, soft masses can be generated at a low messenger scale around $100$ TeV so that soft scalar masses and the gaugino masses are determined by $m^2_i=2N\sum_a C_a (\frac{\alpha_a}{4\pi})^2\frac{F^2}{\Lambda^2}$ and $M_a=N \frac{\alpha_a}{4\pi}\frac{F}{\Lambda}$, with $F<\Lambda^2$, respectively, where $N$ is twice the Dynkin index of the messenger fields, $C_a$ is the quadratic Casimir invariant of group $G_a$, and $F$ is the SUSY breaking F-term. For perturbative unification of gauge couplings, $N\leq 5$ for $\Lambda=100$ TeV.  There is no one-loop trilinear messenger contribution to the A-terms but they are generated by the renormalization group evolution proportional to the gaugino masses. We note that in natural SUSY, one must invoke the SUSY breaking for the first two generation sfermions beyond minimal gauge mediation.
A large splitting between stop/sbottom masses is possible for a large Bino gaugino mass. In this case, a small $\mu$ term, which can be a consequence of natural SUSY, leads to a Higgsino-like MSSM lightest SUSY particle (LSP). On the other hand, if the stop/sbottom mass splitting is small, a Bino gaugino can be the LSP. We assume that the MSSM LSP is long-lived such that it decays outside the detector. This is the case when the gravitino mass is of sub keV \cite{gaugemediation}. Then, as will be discussed later, we can apply the bounds from direct production of sbottoms with the analysis of missing transverse energy plus b-jets.

\subsection{MSSM Higgs and the stop/sbottom ratio}

As noted by many authors a large Higgs mass consistent with current experimental hints is challenging to accommodate
in a minimal SUSY scenario. Working in the decoupling limit and lifting the Higgs mass through one loop stop corrections, one has the relationship \cite{Okada:1990vk,Ellis:1990nz,Haber:1990aw,Martin:1997ns}
\begin{align}\label{Higgsmass}
m_h^2 &= m_Z^2 \cos^2 (2 \, \beta) + \frac{3}{4 \, \pi^2} \, \sin^2 \, \beta \, y_t^2 \, \left[m_t^2 \, \log \left(\frac{m_{\tilde{t}_1} \, m_{\tilde{t}_2}}{m_t^2} \right) + c_{\tilde{t}}^2 \, s_{\tilde{t}}^2 (m_{\tilde{t}_2}^2 - m_{\tilde{t}_1}^2) \, \log \left(m_{\tilde{t}_2}^2/m_{\tilde{t}_1}^2\right) \right.  \\
 &\hspace{6cm} \left.{} \hspace{-0.5cm}+ \frac{c_{\tilde{t}}^4 \, s_{\tilde{t}}^4}{m_t^2} \left((m_{\tilde{t}_2}^2 - m_{\tilde{t}_1}^2)^2 - \frac{1}{2} (m_{\tilde{t}_2}^4 - m_{\tilde{t}_1}^4) \, \log \left(m_{\tilde{t}_2}^2/m_{\tilde{t}_1}^2\right)\right) \right]. \nn
\end{align}

\begin{figure}[h!]
\includegraphics[scale=0.63]{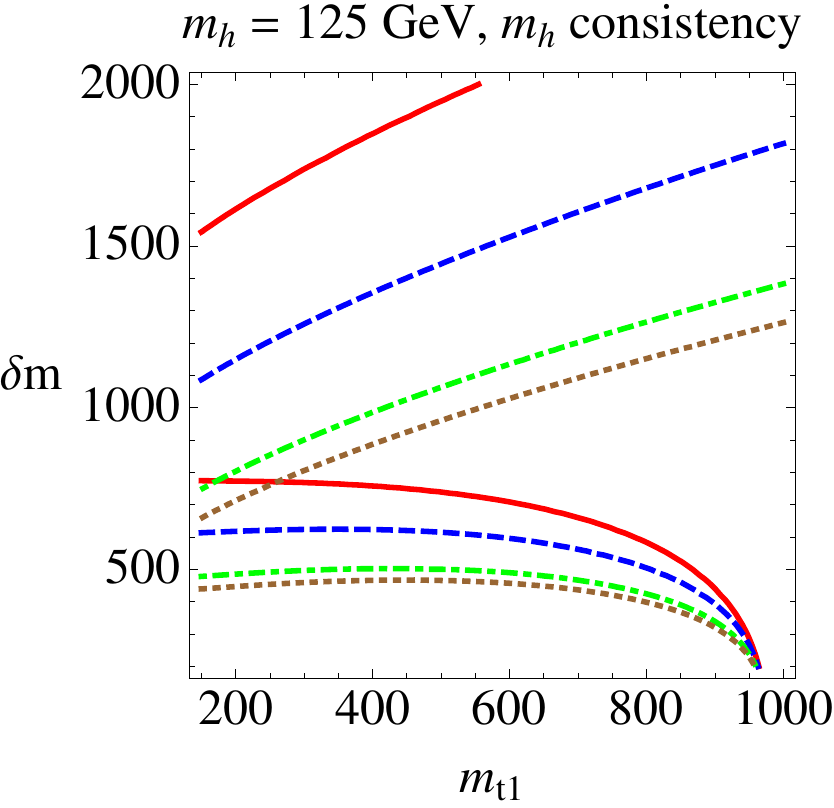}
\includegraphics[scale=0.63]{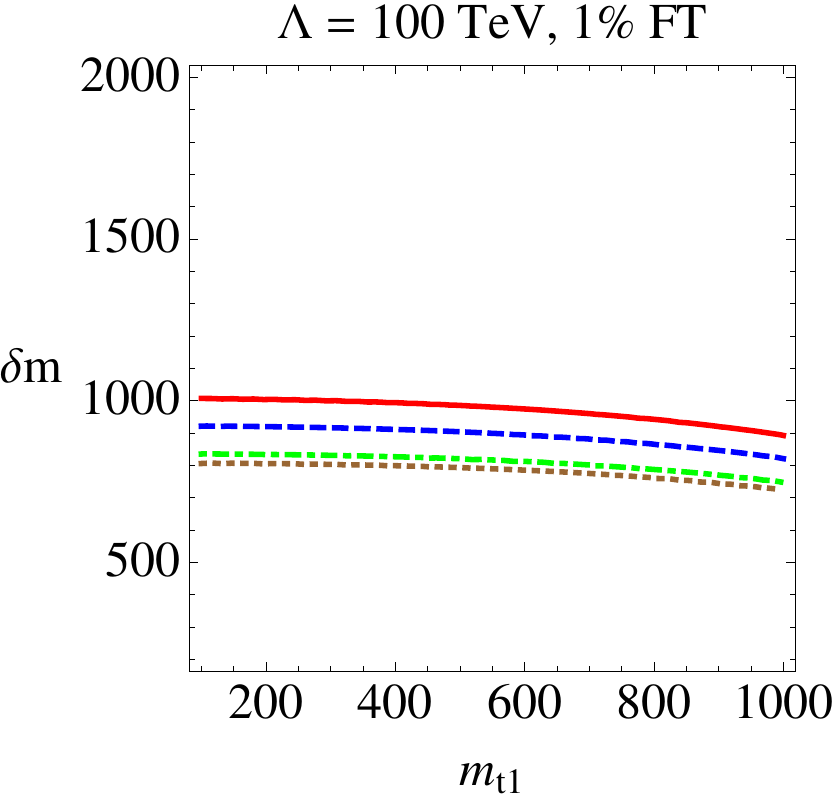}
\includegraphics[scale=0.59]{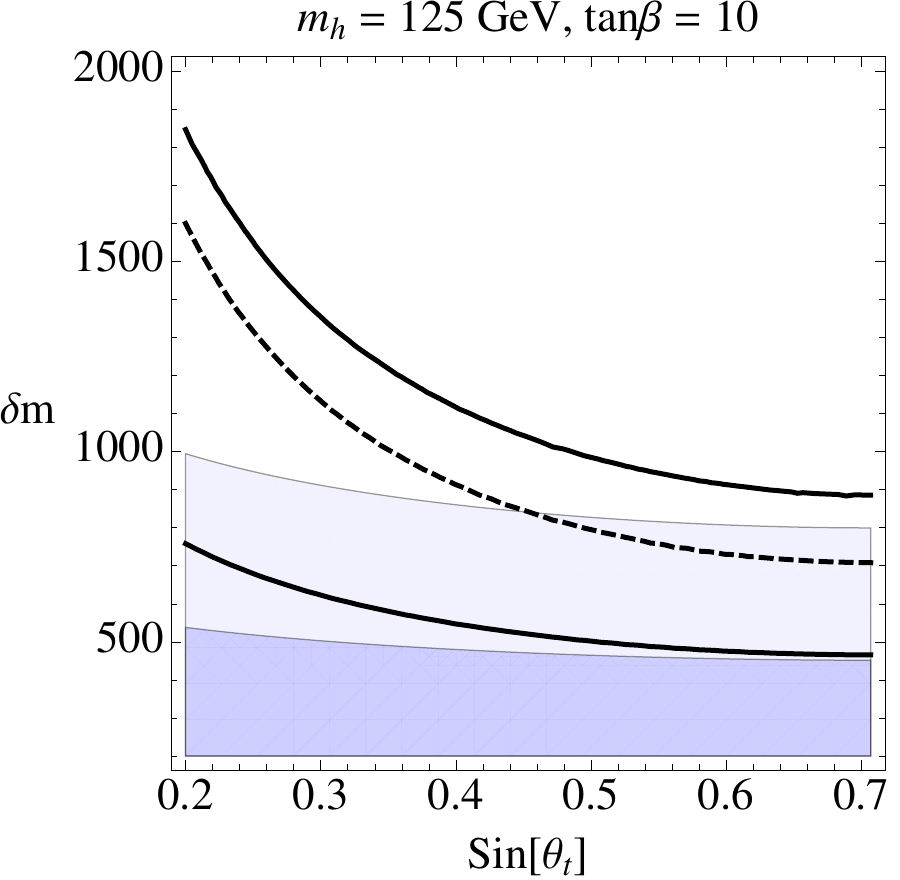}
\caption{The figures illustrate the required stop mass and mass differences to obtain a $125 \, {\rm GeV}$ Higgs for various parameter values (left), the restriction of less than $1\%$ fine tuning on the Z mass according to the defined measure with $\Lambda = 100$ TeV (middle) and in the right figure we show the interplay of the fine tuning, Higgs mass and $\sin \theta_t$ dependence.
As in Fig.~(1), the red solid line corresponds to
$(\tan \beta, \sin \theta_t) = (10,0.2)$, 
the blue dashed line is $(10,0.3)$, the green dot-dashed line is $(10,0.5)$ and the brown dotted line is $(10,1/\sqrt{2})$,
the parameter space below the corresponding line in the middle plot has the defined fine tuning measure $\lesssim 1 \%$.  In the rightmost figure the black solid line is $m_h = 125 \, {\rm GeV}$ using Eqn.~(\ref{Higgsmass}) with $\mu = 400\,{\rm GeV}, m_{\tilde{t}_1} = 400 \, {\rm GeV}$, below the black dashed line is the region of parameter space where the Colour and Charge preserving vacuum condition is satisfied, and the shaded regions are the $1\%$ (lighter shaded region) an $5\%$ fine tuning regions  (darker shaded region).}
\label{initialtheory}
\end{figure}

It is instructive to consider the interplay of imposing that the Higgs mass is lifted by stops and the fine tuning constraint. This illustrates the experimentally derived tension built into a natural SUSY sfermion spectrum.
We plot this relation in Fig.~(\ref{initialtheory}) where we treat the stop masses and the stop mixing angle as free parameters.
As $\tan\beta$ becomes smaller, the fine-tuning measure at given stop masses scales down mildly due to a smaller top Yukawa coupling. However, due to the Higgs mass constraint, the smaller $\tan\beta$, the larger stop masses we need for the Higgs mass, in turn leading to a more fine-tuned situation. As $\Lambda$ is taken to be larger, the fine tuning measure logarithmically scales to require less mass splitting. 
The parameter space most consistent with these constraints is the scenario where $\tan \beta$ takes a moderate value greater than about 5 and stops masses are nearly degenerate.
These general considerations support the point that sbottom exclusions strongly drive stop exclusions though EWPD constraints and $\rm SU_L(2)$ preserving SUSY soft masses quite generally;
a split stop spectrum with $m_{\tilde{t}_1} \gg m_{\tilde{t}_2}$ is disfavoured.  Also note that considering the condition of a colour and charge preserving vacuum further constraints
the parameter space consistent with a $\sim 125 \, {\rm GeV}$ Higgs mass. Insisting on absolute stability when considering this constraint in the minimal MSSM, and attempting to minimize fine tuning to the level of $5 \, \%$ (as shown in Fig.~(\ref{initialtheory},\,right)), supports a maximal mixing scenario with small mass splitting amongst the stop states.\footnote{We have checked the effects of two loop corrections in the case of degenerate spectra and such corrections do not significantly effect the argument for the sub TeV masses we are interested in excluding or discovering.}
 
One could avoid the constraints associated with raising the Higgs mass through a non-minimal scenario such as the NMSSM \cite{nmssm,Hall:2011aa,King:2012is,Ellwanger:2012ke,gnmssm,effmssm}.
So long as the mechanism invoked to raise the Higgs mass does not violate the effective symmetries that are known to be present at
the weak scale, any number of mechanisms can be invoked to raise the Higgs mass. However, even if one remains agnostic about the actual mechanism by which the Higgs mass is raised, 
one still has fine tunings issues that can be minimized by a somewhat degenerate stop spectra with large mixing, and the argument we will advance is still supported by these considerations.
 
\section{Direct Sbottom and stop searches.}\label{direct}

Searches for third generation squarks are a very active area in the SUSY groups at ATLAS and CMS. Several searches with 2011 data have been done, and we briefly review them in this section,
with the aim of illustrating the relative ease of sbottom searches for sub TeV squark masses.

Sbottom searches use a combination of jets, leptons and missing transverse energy plus b-jets. The searches are either gluino-assisted~\cite{gluino-assisted,cms3rd} or based on direct production~\cite{direct-sbottom,cms3rd}. Generally speaking, searches for squarks depend strongly on the gluino mass. The limits from gluino assisted production are model dependent, and lose validity if the gluino becomes so heavy that pair production becomes negligible. For first and second generation squarks, and at moderate values of the gluino mass, t-channel exchange of a gluino is the dominant production mechanism. Hence, setting limits on a simplified model without gluinos, effectively decoupling this particle from the production, leads to weaker bounds than the light squark-gluino searches, see for example Ref.~\cite{multijetsATLAS,multijetsCMS}. With the full 2011 dataset, the bound on the first two generation squarks, independently of the gluino mass is above 1 TeV~\footnote{However, one should keep in mind that the simplified model analysis for multijets and $E_T \! \! \! \! \! \! \! /$ \, is summing over six degenerate squarks to impose a mass bound.}.

With third generation squarks, the t-channel gluino diagram is absent, or very suppressed by PDFs, but searches do still depend on the gluino, if the gluino pair production is significant, and gluinos decay to stops and sbottoms~\cite{gluino-assisted}. Gluino masses are also limited by naturalness, due to their naturalness constraints on Eqn.~(10) at two loops. The gluino is still expected to be in the mass range $m_{\tilde{g}} \sim 1 \, {\rm TeV}$ in a natural SUSY scenario.
The overall production cross sections for sbottoms and stops are numerically very similar for typical natural SUSY third generation squarks.
Although the hadro-production of these states are differentiated by the existence of the $b\bar{b} \rightarrow \tilde{b} \, \tilde{b}$ process with
a t-channel gluino exchange, for gluinos $\gtrsim 1 \, {\rm TeV}$ and  stop and sbottom masses $\lesssim 1 \, {\rm TeV}$, that are of interest in natural SUSY spectra, the impact of this
extra process is $< 1\, \%$ of the overall rate \cite{Beenakker:2010nq}. 

\begin{figure}[h!]
\centering
\includegraphics[scale=0.45]{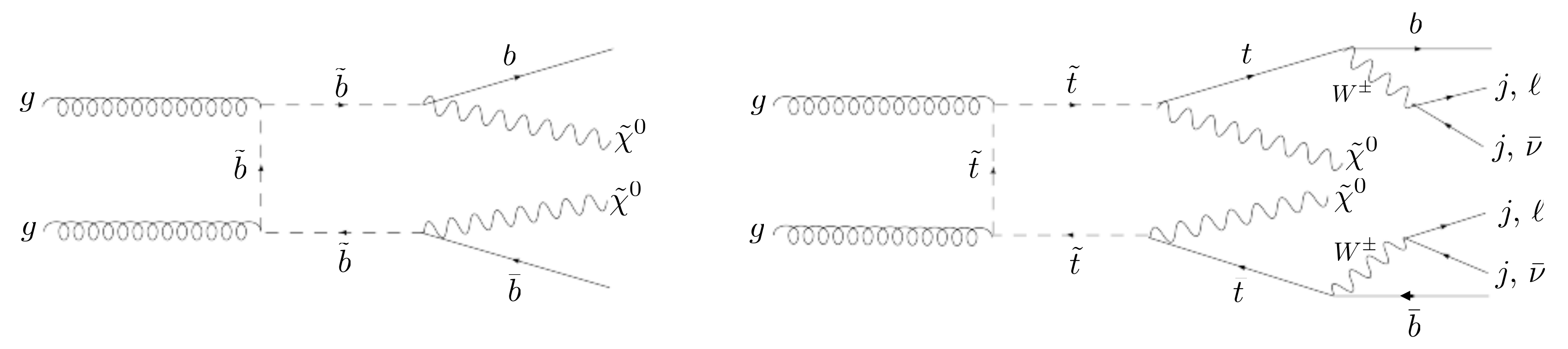}
\caption{Direct production and decay of $\tilde{b} \, \tilde{b}$ and $\tilde{t} \, \tilde{t}$.}
\label{Bstopsbottom}
\end{figure}

\begin{figure}[h!]
\centering
\includegraphics[scale=0.28]{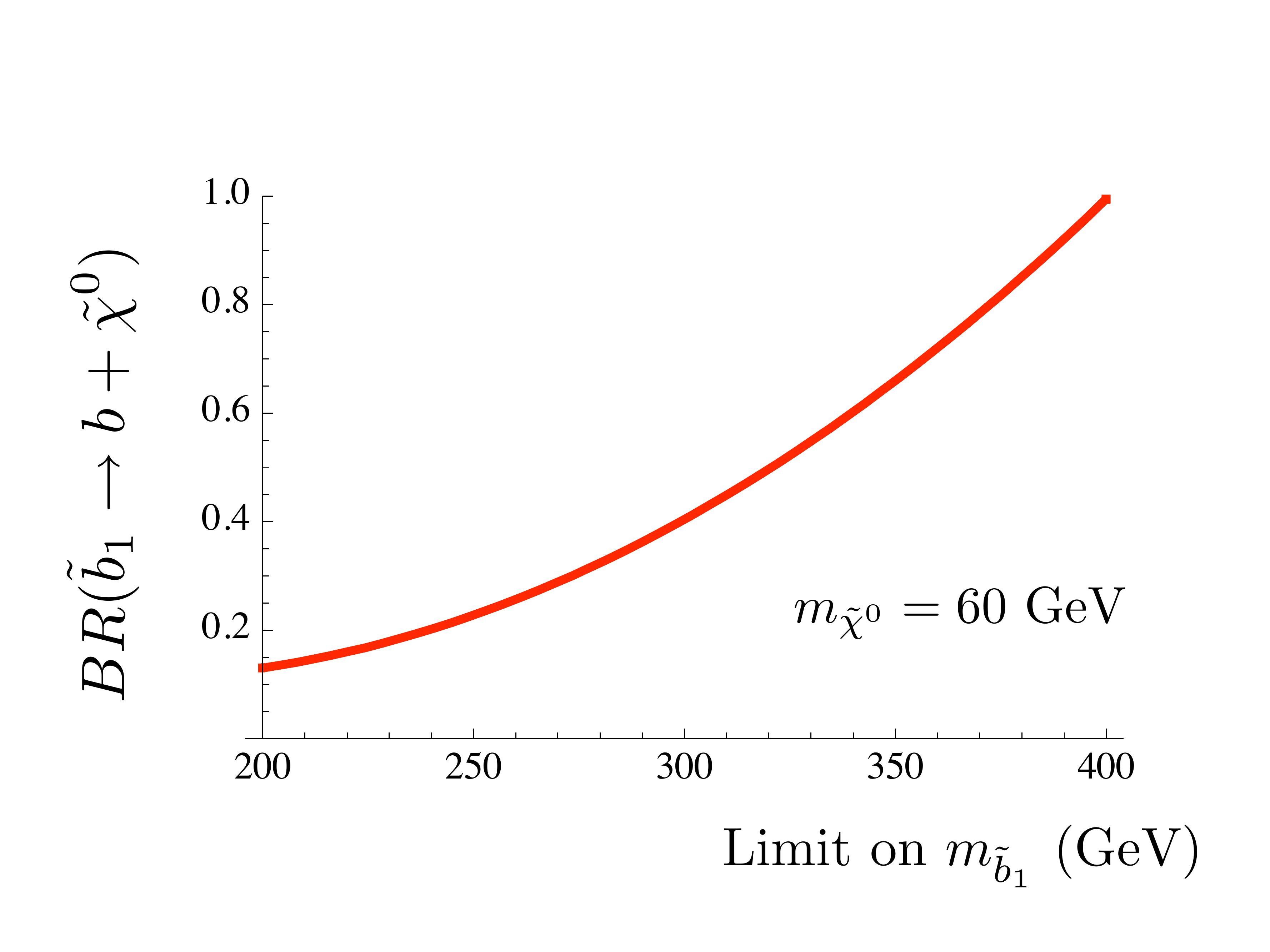}
\caption{Expected limits on sbottom masses as a function of the $\tilde{\chi}^\pm$ branching ratio.}
\label{BR-sbottom}
\end{figure}

Direct production searches are the key ingredient to set a bound on stops and sbottoms which have such similar cross sections for coincident mass scales. 
Robust bounds from direct production of sbottoms can be obtained using triggers on final states with 2 b-jets and $E_T \! \! \! \! \! \! \! /$ \,. A search of this form is interpreted in the context of a simplified model with two parameters (lightest sbottom and LSP masses), assuming ${\rm BR}(\tilde{b}_1\to b$ LSP)=1. With these assumptions, at $2.05 fb^{-1}$ one can exclude sbottom masses up to 390 GeV with an LSP below 60 GeV. The search is most sensitive for $\Delta m=m_{\tilde{b}}-m_{LSP}> 130 $ GeV. See Ref.~\cite{direct-sbottom} for details. We show typical Feynman diagrams appropriate for searches of this form (when decays to a neutralino ($\tilde{\chi}^0$) are kinematically accessible) for the production and decay of stops and sbottoms in Fig.~({\ref{Bstopsbottom}}).
The limit on the lightest sbottom depends on the $\tilde{b}_L$ and $\tilde{b}_R$ admixture in the mass eigenstate and on the nature of the LSP.  If $\tilde{b}_1$ has a large component of $\tilde{b}_L$, the sbottom would also decay to charginos ($\tilde{\chi}^\pm$), hence reducing the BR to the LSP-- provided there is a large mass gap between $\tilde{\chi}^\pm$ and the LSP.  In Fig.~(\ref{BR-sbottom}), we quantify how the limit on the sbottom mass depends on the BR to the LSP, assuming a bound on the sbottom mass of 400 GeV with LHC at 7 TeV. We kept the mass of the $\tilde{\chi}^0$ fixed to 60 GeV, and the separation $\Delta m>$ 130 GeV. The current search is then sensitive to sbottom masses in the 200-400 GeV range, or ${\rm BR}(\tilde{b}_1\to b$ LSP)$>$ 0.13.

Using this approach direct sbottom searches are rather inclusive and range over many SUSY scenarios, however direct stop searches using this approach are not sensitive to date to the expected stop production cross section \cite{Abazov:2009ps,Aaltonen:2011rr,Aaltonen:2011na,Aad:2011wc}. The reason is essentially that the $t \bar{t}$+jets backgrounds are challenging, even with the additional handle on $E_T \! \! \! \! \! \! \! /$ \, and the tagging efficiencies and signal isolate is not sensitive enough. The searches that have been performed that quote stop mass bounds are limited to date, with two kinds of searches available. One assumes gluino assisted production, hence leading to no bound on the stop mass unless a gluino mass measurement is achieved~\cite{gluino-assisted}. The other stop search focuses on a very specific scenario within GMSB, with a stop decaying into a b-jet and $\tilde{\chi}^\pm$ or, if kinematically allowed, into a top and $\tilde{\chi}^0$. The $\tilde{\chi}^0$ is the NLSP, and decays to a Z and gravitino. The search is then based on a final state with two jets (where at least one is tagged as a b-jet), leptonic Z and $E_T \! \! \! \! \! \! \! /$ \,. With these assumptions, one can impose bounds on the $(\tilde{t}, \tilde{\chi}^0)$ mass parameter space. See Ref.~\cite{direct-stop} for details.

\subsection{Reach on stop/sbottom masses with 2012 data}
\begin{figure}[h!]
\centering
\includegraphics[scale=0.32]{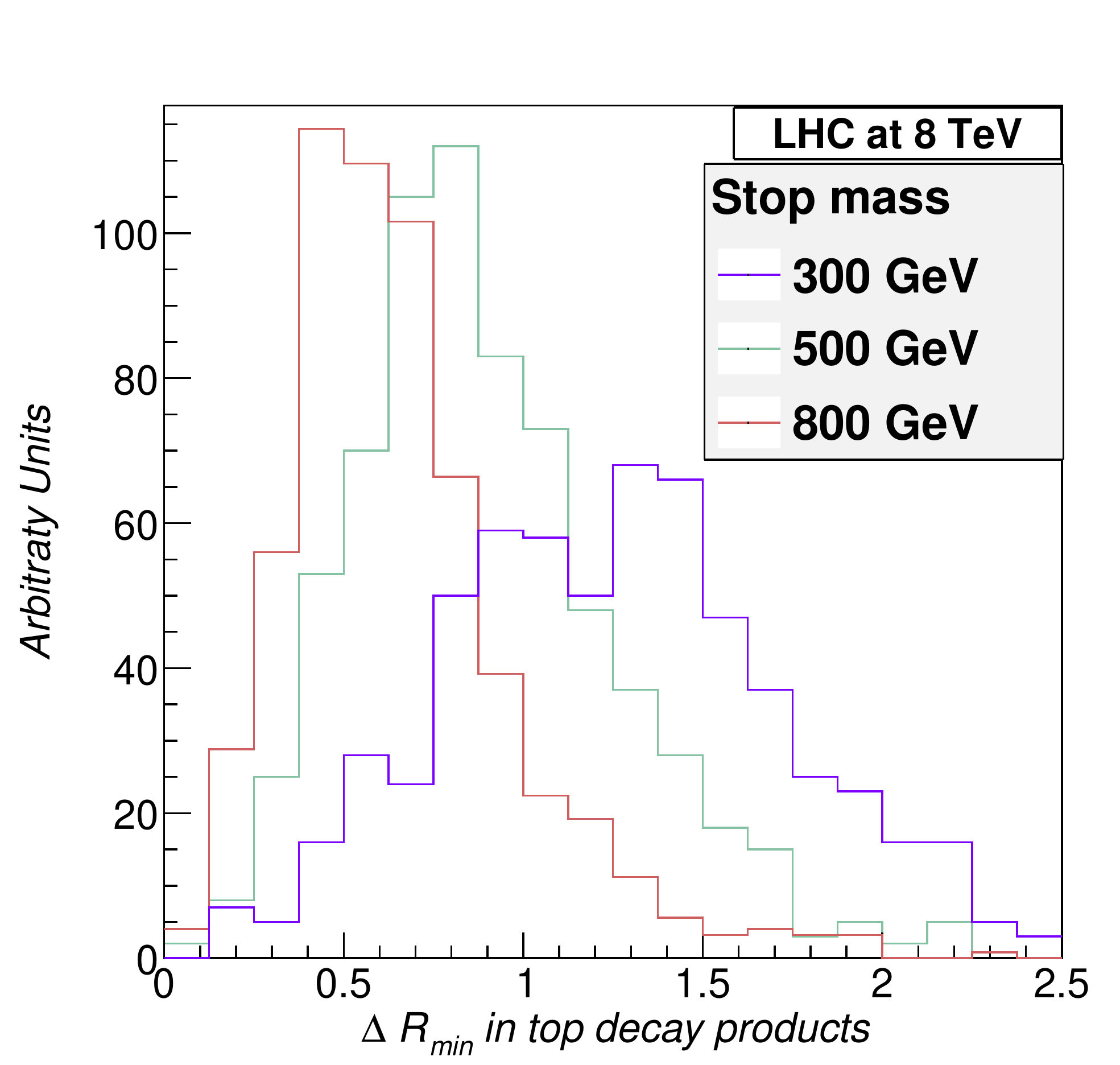}
\caption{Minimum separation in $R$ space between two jets of the top decay coming from a stop.}
\label{topdR}
\end{figure}

Direct stop searches  are also limited by combinatorics and decay topology. Multi-top final states are busy signatures, and new sources of missing energy just add to the complexity of the event. Moreover, at high stop mass, the top becomes boosted, and its decay products tend to merge. In W leptonic decays, isolation criteria would fail to keep the event, and in the W hadronic channel, the jets would tend to merge, tampering with reconstruction and also with b-tagging procedures. To illustrate this, in Fig.~(\ref{topdR}) we plot the minimum separation in $R$ space between two jets from the top decay, when the top comes from a stop of masses 300 to 800 GeV. 

The plots have been generated with Madgraph5~\cite{Alwall:2011uj} for LHC at 8 TeV, with parton level cuts of 20 GeV for all the jets. The jets are parton level objects, where a smearing in energy and momentum has been applied, but no hadronization. The $\Delta R$ we plot is then the separation of the partons: it is an optimistic view of the issue of merging, as hadronization and parton showering, plus clustering, would worsen the plot.  Note that at $m_{\tilde{t}}\simeq$ 500 GeV, the merging becomes sizable.  

The issue of merging is addressed with boosted top techniques. Unfortunately, these techniques have a limited range of efficiencies, as a function of top $p_T$. The efficiency reaches 45\% for a range of $p_T>$600 GeV. Stops of mass 500 GeV would lead to 3\% of the tops in that range, whereas for stops of mass 1 TeV, 30\% of the tops would satisfy this bound on $p_T$~\cite{boosted-tops}. Since with 2012 data due to PDF effects and limited statistics at large invariant mases, one could expect reaching production cross section limits on third generation for masses in the few hundreds of GeVs, an efficiency of few percent seems rather discouraging.

\begin{figure}[h!]
\centering
\includegraphics[scale=0.32]{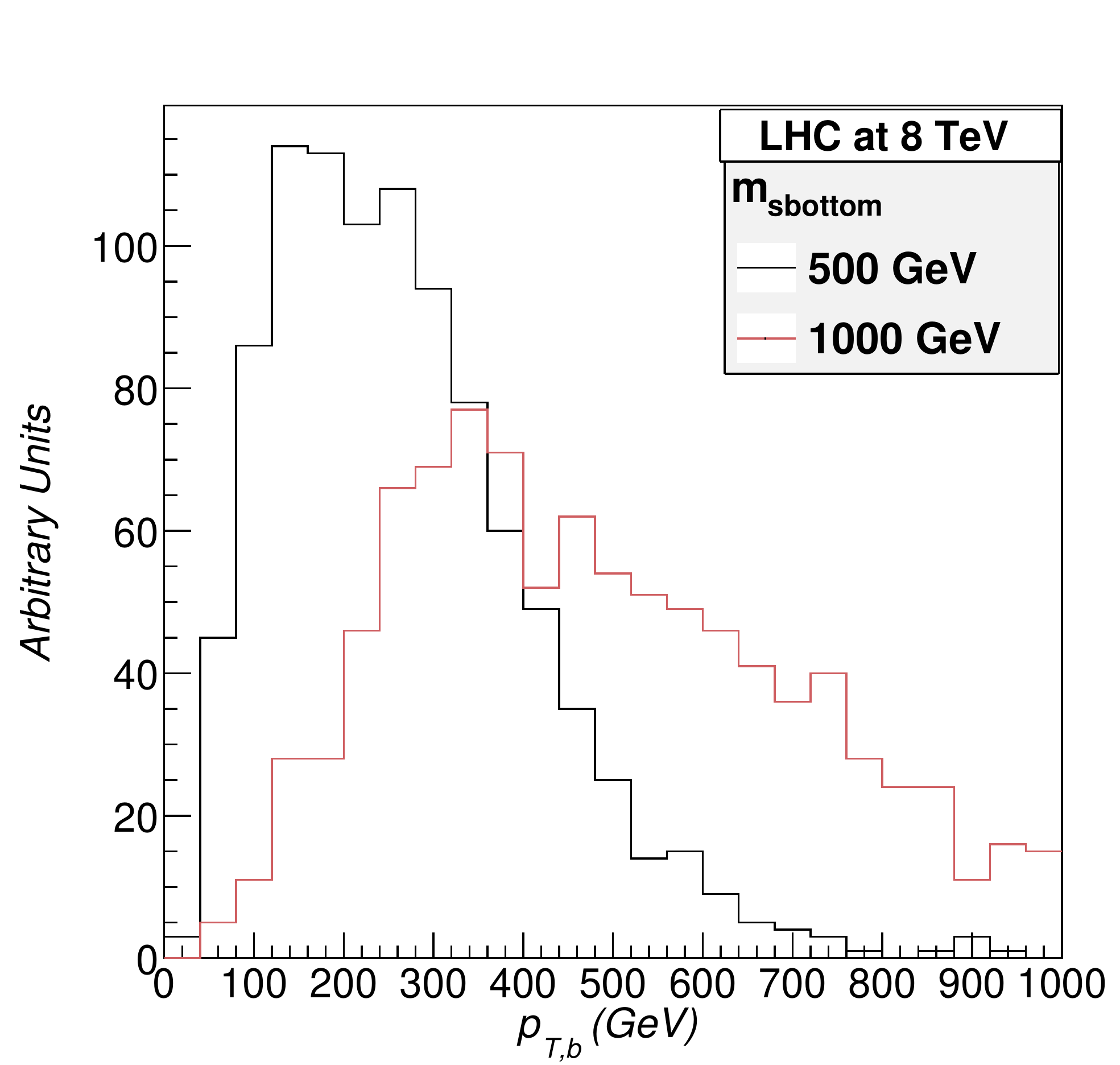}
\caption{$p_T$ distribution of the b-jet coming from the sbottom decays. The neutralino mass is fixed to 100 GeV.}
\label{Efficiencies}
\end{figure}

On the other hand, sbottom searches based on b-jets+$E_T \! \! \! \! \! \! \! /$ \,\, are more easy to scale to higher values of sbottom masses in the interesting range for natural SUSY theories. b-tagging efficiency is relatively stable with transverse momentum ($p_T$), ranging from 60-80\% for $p_T$ in the range below 670 GeV~\cite{CMSnote}. Commissioning for b-tagging for the 8 TeV run is not yet public, but we will assume the efficiencies would remain in that range. In Fig.~(\ref{Efficiencies}) we plot the $p_T$ distribution of the b-jet coming from the sbottom decays when the neutralino mass is fixed to 100 GeV. In the range of $p_T$ shown here, we expect that the known b-tagging algorithms would be applicable. 

 One can estimate the reach on sbottom masses in the search described in Ref.~\cite{direct-sbottom} with the 2011 full dataset, assuming there are no improvements in the efficiency so that the limit on the total cross section just scales with $\sqrt{ {\cal L}}$. The result of this exercise leads to an increase from 390 GeV to 420 GeV. The running of 2012 at 8 TeV will further increase the sbottom bound, as that pair production cross section from 7 to 8 TeV increase by a factor 2(3)  for a 500 GeV(1 TeV) sbottom. Assuming $20 \, fb^{-1}$ of data per experiment, and estimating that the background fraction would not increase significantly from the 7 TeV run, one can forcast a reach of $\lesssim 800 \, {\rm GeV}$. An increase in efficiency of the analysis would probably be required to access sbottom masses above TeV with the 2012 run. 

To summarize the experimental situation we show in Table~\ref{table} the efficiencies for basic cuts in stop and sbottom searches, for a range of masses up to 1 TeV. 

\begin{table}
        \begin{center}
\begin{tabular}{|p{2.2cm}|| p{4.7cm} |p{4.7cm} |p{4.7cm}|}\hline 
 ~ & {\bf Sbottom} & {\bf Stop, non boosted} & {\bf Stop, boosted}   \\   &  For $p_T^b < 670$ GeV &  SM $t\, \bar{t}$ similar.   & Top-tagging eff.$\gtrsim$40\% \\
  & $\epsilon_{b,tag}\simeq 60-80$\%  \cite{CMSnote} & Considering only lepton & if  $p_T^t \in[600,1600]$  GeV \cite{boosted-tops} 
  \\   & $\epsilon_{mistag}\simeq 1-10 \%$ ~\cite{CMSnote}. & isolation criteria $\Delta R> 0.7$.  & \\ \hline & $\hspace{1.5cm} \epsilon_{p_T^b<670}$ & $\hspace{1.5cm}  \epsilon_{\Delta R > 0.7}$ & $\hspace{1.5cm} \epsilon_{p_T^{top}>600}$ \\ \hline $<$ 300 GeV & 1 &  $>$ 0.50 & $<$ 0.01\\  300-700 GeV &  $\simeq$ 1 & 0.50-0.25 & 0.01-0.1\\   700-1000 GeV & $>$0.78 & $<$0.25& 0.1-0.3 \\ \hline
    \end{tabular}
         \end{center}
\caption{Estimated efficiencies $\epsilon_i$ for basic cuts in searches for sbottoms and stops, for LHC at 8 TeV.}
\label{table}
\end{table}

In the first column, we use the information in Ref.~\cite{CMSnote} on b-tagging. In this note from CMS, different b-tagging algorithms are compared in the 7 TeV run (no similar study is available for the 8 TeV run). Those algorithms achieve a 60-80\% tagging efficiency (in $t\bar{t}$ samples) with a mistag rate in the range of few percent. Only the region of b-jet $p_T< $ 670 GeV contains enough statistics to study the algorithms. One could worry that, for high sbottoms masses, the b-jet could be very boosted, beyond what is described in the  CMS note.  We take a conservative approach and ask for a cut on  b-jet $p_T<$ 670 GeV, the reach of the CMS study. Even at 1 TeV, the cut only reduces $\sim$20\% of the signal. Assuming the commissioning of b-tagging at 8 TeV is as efficient as in the 7 TeV run, the issue of tagging b-jets coming from the decay of sbottoms up to TeV would have a stable efficiency, in the range 50-80\% per b-tag. 

In the second column of Table.~\ref{table}, we discuss the stop searches without boosted techniques. There is no public experimental study in this regime, and we cannot estimate the efficiency of the cuts required to reduce the backgrounds to acceptable levels. We quote a cut on lepton isolation which will be basic in all leptonic studies: top backgrounds are a major issue in these searches, and would require b-tagging and possibly leptonic decays, which require some isolation cut. Usual isolation cuts are $\Delta R>0.7$, a criteria which is harder to satisfy as the stop mass increases~\footnote{Note that a $p_T$ dependent cut could improve the situation~\cite{deltaRpt}, but to our knowledge there is no experimental study on the efficiency of varying the $\Delta R$ cut based on the $p_T$ of the objects.}. Indeed, whereas for masses below 300 GeV, the cut is more than 50\% efficient, for masses above 700 GeV, only 25\% of the events would pass isolation requirements. Those numbers correspond to one of the tops decay products. Asking for both tops products passing the isolation cuts would correspond to an efficiency $\epsilon_{\Delta R}^2$.

Finally, in the last column we discuss stop searches using boosted techniques. Those are very promising for large stop masses~\cite{boosted-tops}. There are many proposals of top-tagging techniques, and we are going to focus on the Johns Hopkins algorithm~\cite{JHtop}~\footnote{Note that for low top $p_T$, the HEP algorithm~\cite{HEPtop} may be more efficient than the Johns Hopkins~\cite{JHtop}.}. They reach an efficiency in the 40-50\% level for tops with $p_T>$ 600 GeV. Below this value, the efficiency degrades very fast. We then quote what is the efficiency for one of the tops from the stop decays to pass this cut, for several values of stop masses. Again, asking for both tops in that range would lead to an efficiency of $\epsilon_{p_T^t>600}^2$. Note that for the range of masses 300-700 GeV, the efficiency is very low ($10^{-5}-10^{-3}$ for two top-tags), as compared with the b-tagging efficiencies from  sbottom searches ($\gtrsim 0.3$ for two b-tags). Top-tagging techniques could improve dramatically this year, but they should do so by orders of magnitude to reach the sensitivity of sbottoms searches at the same mass point.

These considerations show that in the interesting mass range for stops in natural SUSY, $m_{\tilde{t}_1}, m_{\tilde{t}_2} \lesssim 1 \, {\rm TeV}$, the related bounds on sbottoms discussed are likely to be the most sensitive experimental probe
for large regions of the $(m_{\tilde{t}_1},m_{\tilde{t}_2}, \theta_{\tilde{t}})$ space, and hopes for natural SUSY can hit sbottom in the 2012 run.

\section{Natural MSSM Higgs and the stop/sbottom mass limits.}\label{interplay}

\begin{figure}[h!]
\centering
\includegraphics[scale=0.22]{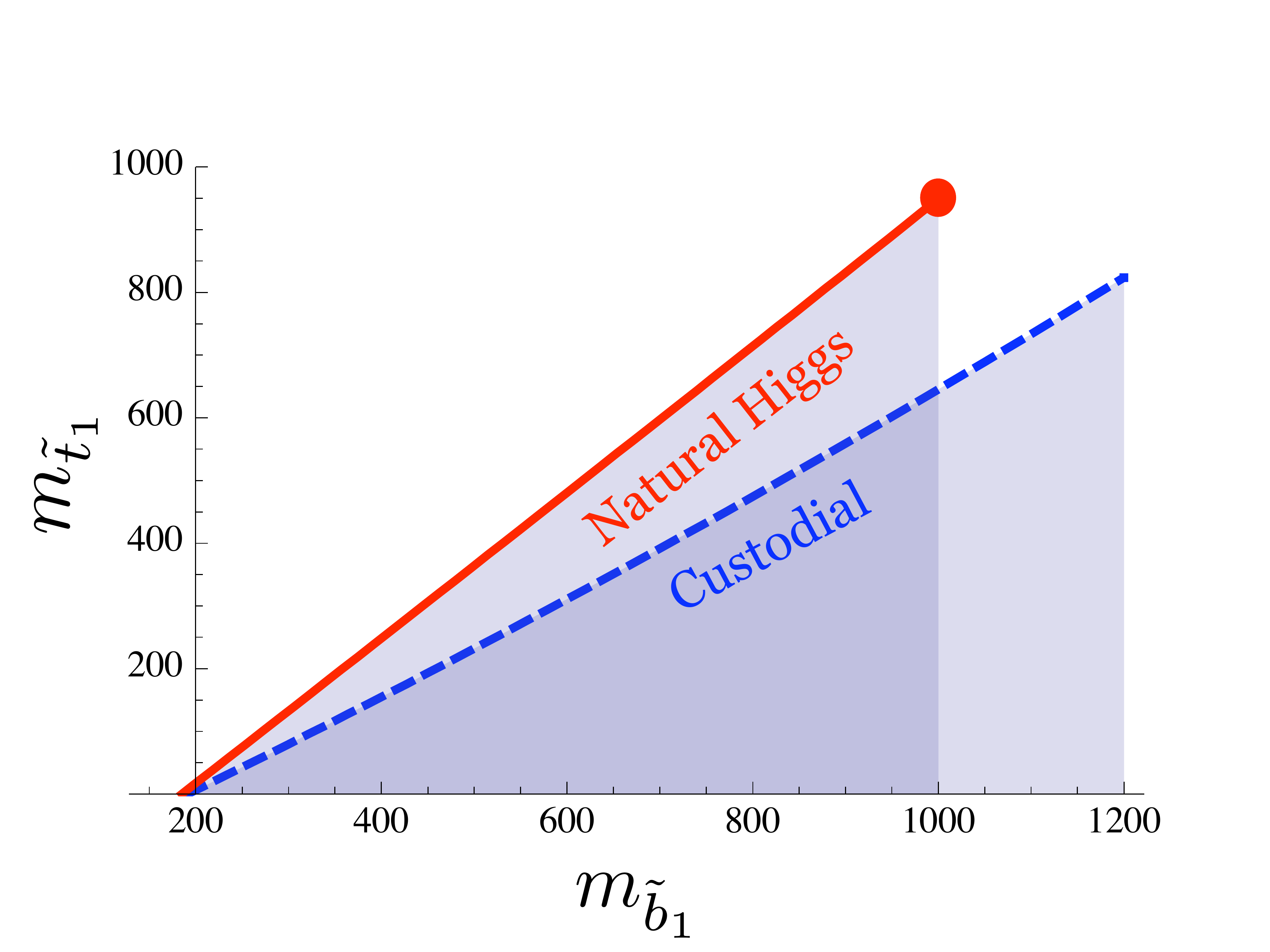}
\includegraphics[scale=0.23]{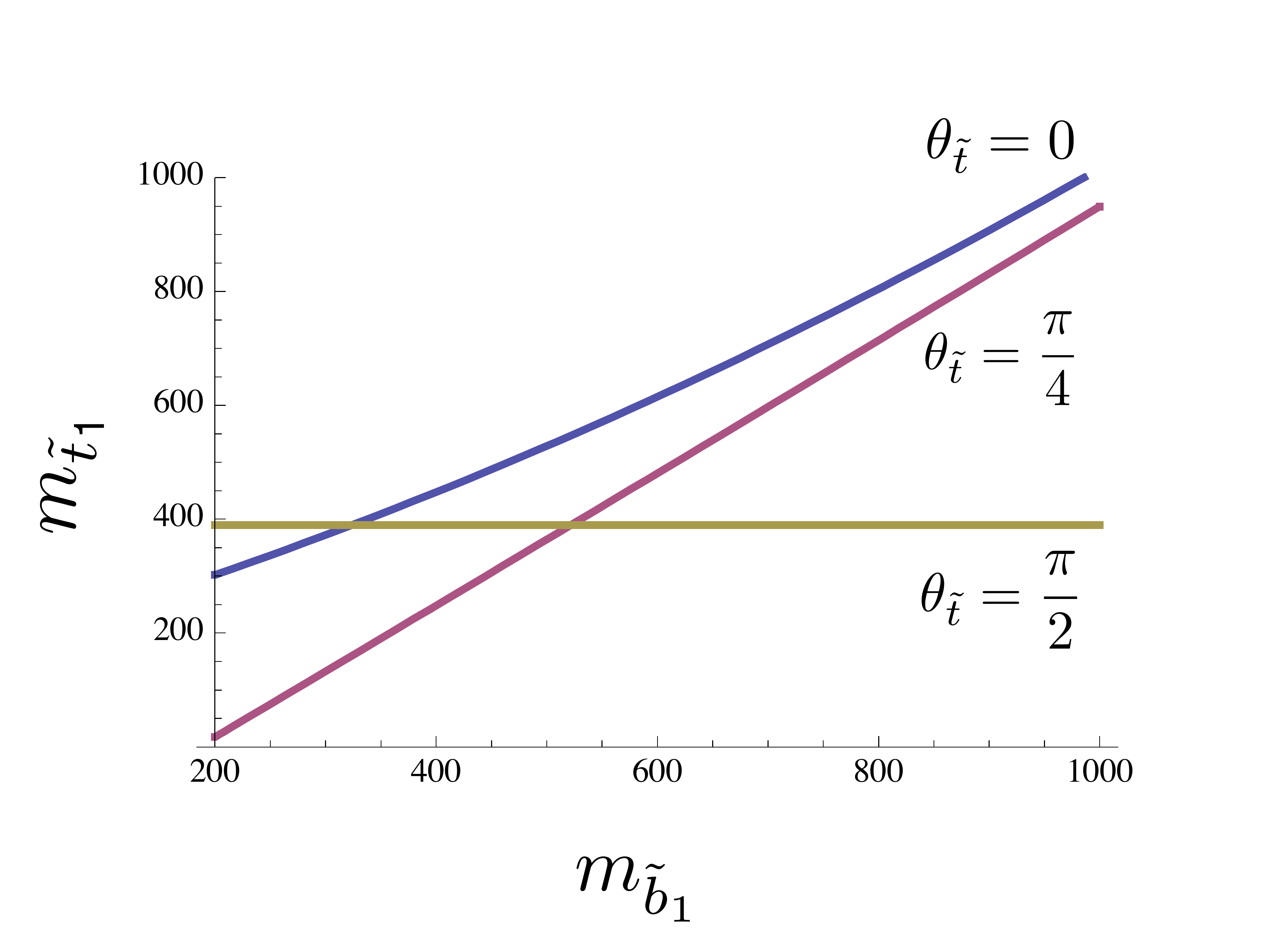}
\caption{Left:The sbottom bound versus the stop bound using restrictions from custodial violations (blue line) and a natural MSSM Higgs, with a finetuning at the level of 1\%. Maximal mixing case. 
Right:The sbottom bound versus stop bound imposing constaints from violations of custodial symmetry for different stop mixing angles.}
\label{cust-tot}
\end{figure}

In this section, we explore in more detail the interplay of the three sources of constraints on stops discussed in Sec.~\ref{theory}: $\rm SU_C(2)$ violation, naturalness and a MSSM Higgs. $\rm SU_C(2)$ violation bounds from $\Delta \rho$ already relates the stop and sbottom sectors, but accommodating a natural MSSM Higgs at $m_h \sim125 \, {\rm GeV}$ in the theory adds an even stronger correlation between the two sectors. We illustrate this point in Fig.~(\ref{cust-tot}), where we plot the bound on the lightest stop derived from a bound on the lightest sbottom. The blue line corresponds to imposing 
$\rm SU_C(2)$ constraints, whereas the red line corresponds to adding the constraints of accommodating a natural Higgs in the MSSM. We chose the maximal stop mixing case, and a value of finetuning of 1\%. The end-point of the red line corresponds to the situation where no solutions with less than 1\% fine-tuning are obtained.

One may wonder how those constraints vary with the stop mixing angle, as we know the custodial constraints are weakened when the lightest stop is purely right-handed. In Fig.~(\ref{cust-tot}), we show the effect of this variation, when constraints from violations of custodial symmetry are applied. If the lightest stop is purely right-handed, there is no correlation between a bound on sbottoms and the lightest stop. This bound would only be correlated with the heaviest stop. But if the lightest stop has any admixture of left-handed stop, improvements on the sbottom bounds lead to a push of the lightest stop mass. If we also imposed a constraint on naturalness, or the MSSM Higgs, even the case of the light right-handed stop becomes correlated with sbottom searches as we have discussed, as a nearly degenerate spectra is selected for. Note that mixing angles are not renormalization group invariant. Invoking particular mixing angles to disassociate the stop and sbottom sectors requires further tuning of parameters.


\section{Conclusions}

The testing ground for natural SUSY is widely considered to be direct production in  the stop sector, but direct access to this sector is extremely experimentally challenging.

In this paper, we have exploited the minimal consistency constraints associated with experimentally motivated limits of custodial symmetry violation to link the stops-- and the issue of fine-tuning-- to the comparatively cleaner and more promising searches for sbottoms.  For example, a sbottom bound of 500 GeV translates into a degree of fine-tuning in the theory of at least 5\%, whereas setting a sbottom bound on the 1.5 TeV range, pushes the fine-tuning below the 1\% level.
Even if EWPD is ignored, a strong relationship between the mass scale of sbottoms and stops follows from only assuming that soft SUSY masses are $\rm SU_L(2)$ invariant.

These links between direct sbottom searches and the stop parameter space of interest in natural SUSY scenarios are made even stronger
when an MSSM Higgs in the 125 GeV region is itself associated with the stop spectrum. Although unknown mixing angles mean that a mapping of the excluded sbottom space is related
to a range of stop masses, the relationship between the sectors is strong enough that sbottom searches
can be reasonable expected to largely drive the exclusion of the stop parameter space in the 2012 run. Hopes for a natural SUSY may thus sbottom out experimentally. Conversely,
if weak scale natural SUSY reveals itself in the 2012 run, a sbottom up discovery of natural SUSY is favoured by these same arguments. 

{\raggedleft{\large {\it ``I'll speak in a monstrous little voice."}}\\
\raggedleft{{(s?)bottom -- A Midsummers Night's Dream}\\
\raggedleft{ {\it Act I, Scene ii}}}

\subsection*{Acknowledgments}
\raggedright
MT thanks James Wells for a helpful conversation concerning radical naturalists. We thank G. Salam for helpful comments on top and bottom tagging.

\end{document}